# Cooperative Routing for Wireless Networks using Mutual-Information Accumulation

Stark C. Draper, *Member, IEEE,* Lingjia Liu, *Member, IEEE,* Andreas F. Molisch, *Fellow, IEEE,*
Jonathan S. Yedidia, *Member, IEEE*

*Abstract*—Cooperation between the nodes of wireless multihop networks can increase communication reliability, reduce energy consumption, and decrease latency. The possible improvements are even greater when nodes perform mutual information accumulation using rateless codes. In this paper, we investigate routing problems in such networks. Given a network, a source, and a destination, our objective is to minimize end-to-end transmission delay under energy and bandwidth constraints. We provide an algorithm that determines which nodes should participate in forwarding the message and what resources (time, energy, bandwidth) should be allocated to each.

Our approach factors into two sub-problems, each of which can be solved efficiently. For any transmission order we show that solving for the optimum resource allocation can be formulated as a linear programming problem. We then show that the transmission order can be improved systematically by swapping nodes based on the solution of the linear program. Solving a sequence of linear programs leads to a locally optimal solution in a very efficient manner. In comparison to the proposed cooperative routing solution, it is observed that conventional shortest path multihop routing typically incurs additional delays and energy expenditures on the order of 70%.

Our first algorithm is centralized, assuming that routing computations can be done at a central processor with full access to channel state information for the entire system. We also design two distributed routing algorithms that require only local channel state information. We provide simulations showing that for the same networks the distributed algorithms find routes that are only about two to five percent less efficient than the centralized algorithm.

## I. Introduction

Multihop relay networks are one of the most active research topics in wireless communications. The use of relays enables a number of performance improvements. Energy efficiency can be improved since the distances over which each node must transmit are often reduced significantly. Improved robustness to fading and failure of individual nodes results from the increased number of possible transmission paths connecting source and destination, reducing the probability of loss of session connectivity.

The most basic form of relaying consists of routing information along a single path. Data packets are passed from one node to the next in a manner akin to a bucket brigade. For example, this approach underlies the widely used Zigbee standard [1] for low-rate, low-power networking. More sophisticated methods that require tighter synchronization between nodes at the physical and media access control (MAC) layer can lead to much larger performance gains; see, e.g., [2]–[6] and the references therein.

At a high level multihop relaying can be broken down into two distinct sub-problems. The first is the design of physical and MAC layer techniques for relaying information from one set of nodes to the next. The second is routing, i.e., identifying which of the available nodes should participate in the transmission and what system resources (time, energy, bandwidth) should be allocated to each. These two sub-problems are connected. As we see in this paper the physical layer technique employed strongly influences the optimum route.

Most of the routing papers in the literature are based on physical layer techniques that either use virtual beamforming or energy accumulation. In virtual beamforming the amplitude and phases of the signals at transmitting nodes are adjusted to interfere constructively at the receiver [7]–[9]. In energy accumulation multiple transmissions are combined non-coherently by receiving nodes. This is enabled, e.g., through space-time or repetition coding [10], [11], [25]. A different approach based on mutual-information accumulation is proposed in [12], [13].

The difference between energy accumulation and mutual information accumulation is most easily understood from the following example. Consider binary signalling over a pair of independent erasure channels each having erasure probability $p_e$ from two relays to a single receiver. If the two relays use repetition coding, corresponding to energy accumulation, then each symbol will be erased with probability $p_e^2$. Therefore, $1 - p_e^2$ novel parity symbols are received on average per transmission of the two transmitters. On the other hand, if the two transmitters use different codes, the transmissions are independent and on average $2(1 - p_e)$ novel parity symbols (which exceeds $1 - p_e^2$) are received per transmission.

For Gaussian channels (or fading channels with decoder channel state information) at low signal-to-noise ratios (SNRs) energy accumulation is equivalent to mutual-information accumulation because at low SNRs capacity is approximately linear in SNR. However, as SNR increases, mutual-information accumulation gives better results than the either virtual beamforming or energy accumulation. For this reason mutual-information accumulation is the physical-layer technique used in this paper. Mutual information accumulation can be realized through the use of rateless codes of which Fountain and Raptor codes [16]–[18] are two prominent examples.

The primary contributions of the current paper are threefold.

S. C. Draper is with the Dept. of Electrical and Computer Engineering, University of Wisconsin, Madison, WI 53706 (E-mail: sdraper@ece.wisc.edu).

L. Liu is with Dallas Telecomm. R&D Center, Samsung Electronics (E-mail: lliu@sta.samsung.com).

A. F. Molisch is with the Department of Electrical Engineering, University of Southern California, Los Angeles, CA, 90089, USA. (email: Andreas.Molisch@ieee.org).

J. S. Yedidia is with the Mitsubishi Electric Research Laboratories, Cambridge, MA 02139 (Email: yedidia@merl.com).

2- First, we present a mathematical formulation of the routing problem with mutual-information accumulation where the objective is to minimize end-to-end delay under various bandwidth and energy constraints. The cases of energy minimization under end-to-end delay and bandwidth constraints or of bandwidth minimization under end-to-end delay and energy constraints can be treated in a completely analogous manner.
- Second, under the assumption of centrally available and complete channel state information (CSI), we detail an iterative method to optimize the route based on solving a sequence of linear programs. Each linear program solves for the optimal resource allocation for a given route. The resulting allocation is then used to update the route and the method proceeds iteratively. By leveraging our solution to revise the route, the proposed algorithm can find a "good" route very efficiently.
- Finally, taking inspiration from our centralized solution, we provide two distributed algorithms that require only local CSI. Simulations show that the resulting solutions require less than 5% additional energy for the same end-to-end delay as the centralized solution.

To our knowledge, there has been little prior work investigating routing in networks consisting of nodes using mutual-information accumulation. In [12] mutual information accumulation is considered for a single-relay network. Mutual information accumulation is also investigated in [13], but the analysis therein assumes network "flooding", i.e., all nodes transmit all the time; this is not an optimum use of energy. Regarding linear-programming based routing solutions for adhoc networks, in [10], [11] the routing problem is posed as a linear-program, but the physical layer technique assumed is energy accumulation. Furthermore, the outcome of the linear-program is not further explored to improve the selected route. Another heuristic algorithm for routing with energy accumulation was proposed in [25]. In [24] a heuristic algorithm for relaying information with hybrid ARQ (automatic repeat request) with mutual information accumulation *over time* is derived. In contrast to our paper, however, [24] assume that when relay nodes transmit simultaneously, they send out the same signal.

An outline of the paper is as follows. We present the system model in Sec. II. We present and discuss illustrative results in Sec. III. The centralized routing and resource allocation algorithm, and its constituent parts, are developed in Sec. IV. In Sec. V we describe the two distributed algorithms. We provide details of simulation results in Sec. VI and conclude in Sec. VII. Proofs are provided in the appendix.

## II. System model

In this section we present our system model. We consider a uni-cast network consisting of $N+1$ nodes: the source, the destination, and $N-1$ relay nodes. The network's objective is to convey a data packet composed of $B$ bits from source to destination in the minimum time under sum-energy and bandwidth constraints.[1] The relays may participate actively in packet transmission or may remain silent for the duration of communication. Relay nodes operate under a half-duplex constraint: they can either transmit or receive but cannot do both simultaneously. To simplify analysis we assume that a node's only significant energy expenditure lies in transmission; reception, decoding, and re-encoding entail no significant overhead. We note that this assumption can be relaxed within the framework presented.

The $i$th node operates at a fixed transmit power spectral density (PSD) $P_i$ (joules/sec/Hz), uniform across its transmission band. The propagation channel between each pair of nodes is modeled as frequency-flat and block-fading, where the coherence time of the channel is larger than any considered transmission time of the encoded bits. The channel power gain between the $i$th and the $k$th nodes is denoted $h_{i,k}$. Under these assumptions, the spectral efficiency of data transmission from node $i$ to node $k$ can be expressed, following Shannon's classical formula [21], as

$$C_{i,k} = \log_2\left[1 + \frac{h_{i,k}P_iW_i}{N_0W_i}\right] = \log_2\left[1 + \frac{h_{i,k}P_i}{N_0}\right] \text{ bits/sec/Hz}, \quad (1)$$

where $N_0/2$ denotes the PSD of the (white) noise process.

If node $i$ is allocated the time-bandwidth product $A_i$ sec-Hz for transmission, the potential information flow from node $i$ to node $k$ is $A_iC_{i,k}$ bits. Our first assumption is that nodes use codes that are ideal in the sense that they fully capture this potential flow, working at the Shannon limit at any rate. Nodes are further designed to use *independently generated* codes for relaying. This design choice connects to our second assumption which is that, without any rate loss, a receiver can combine information flows from two or more transmitters. If, for example, transmitting nodes $i$ and $j$ are allocated time-bandwidth products $A_i$ and $A_j$, respectively, our two assumptions mean that node $k$ can decode as long as the mutual information accumulated by node $k$ exceeds the message size, i.e.,

$$A_iC_{i,k} + A_jC_{j,k} \geq B. \quad (2)$$

The use of independently-generated codes is crucial to the mutual-information accumulation condition reflected in (2). If the *same* code were used by each transmitter, the receiver would get multiple looks at each codeword symbol. This is "energy-accumulation." By getting looks at different codes (generated from the same $B$ information bits) the receiver accumulates mutual information rather than energy.

Although other implementations are possible, the two assumptions of ideal codes and mutual-information accumulation from multiple streams can most naturally be realized (albeit approximately) through the use of "fountain" (or "rateless") codes [20]. Fountain codes encode information bits into potentially infinite-length codewords; additional parity symbols are sent by the transmitter until the receiver is able to decode. For a discussion of how well fountain codes can fulfill our assumptions, see, e.g., reference [13]. The non-ideal nature of existing implementations of fountain codes can be handled in the optimization framework of this paper without undue trouble. For example, one can incorporate an overhead factor of $(1 + \epsilon)$ into the right-hand side of (2).

While the example of (2) considers only two nodes, in general a receiver will combine receptions from all transmitting nodes to recover the data. The only requirement for decoding is that the total received mutual information, summing over all transmitting nodes, exceeds $B$ bits [13].

---

[1]Multiple messages can be transmitted in parallel over (quasi-) orthogonal channels. See the discussion in [19] and [13].

The network also operates under bandwidth and energy constraints. We study the case where these resources are constrained on a per-node basis and also the case where the constraints are imposed on the sum allocation across nodes. Such constraints involve the $A_i$ and the $A_i P_i$ products. Full details will be provided in Section IV.

## III. Motivation

In this section we illustrate the improvements made possible by combining mutual information accumulation with route optimization for a simple one-dimensional network. This model is amenable to closed-form analysis. We present these results prior to their full derivation in Section IV-E, so that readers can develop a sense of the possible improvement before delving into the full details of the algorithms and analysis.

The one-dimension network we consider consists of $N + 1$ nodes equally-spaced along the line segment $[0, D]$. The source node 0 is located at the origin and the destination node $N$ is located at $D$. The channel power gain between two nodes, $i < j$, is proportional to $(d_{i,j})^{-2} = (N/D)^2 (i - j)^{-2}$. As is fully developed in Section IV-E, under a system-wide sum-bandwidth constraint $W_T$, we can analytically solve for the transmission duration $\tau_c$ achieved by our cooperative protocol.

Consider the case where $P_i = P$ for all $i$. In this case the cooperative strategy that minimizes the transmission duration $\tau_c$ is for the source (node 0) to transmit long enough that node 1 can decode the message and then to stop transmitting. At that point node 1 starts to transmit (since it has received the packet) and its connectivity $C_{1,k} > C_{0,k}$ for $k > 1$ (since $P_i = P$ for all $i$ and $d_{1,k} < d_{0,k}$). Thus it is better to allocate the full system bandwidth to node 1 rather than reserving some so that node 0 can continue to transmit. Subsequent transmissions last until the next node decodes. For example, the transmission from node $i$ lasts until node $i + 1$ decodes. Each transmission is shorter than the previous transmission. This is due to the mutual information already accumulated by nodes further down the chain during earlier nodes' transmissions. The process of "passing on" the information from node to node continues until the destination decodes the packet.

For comparison we also solve for $\tau_{nc}$ the transmission duration achieved by the best non-cooperative scheme where mutual-information accumulation is not performed. In this protocol each node listens only to a *single* transmission. Unlike in the cooperative system, in such a system the optimal route depends on the magnitude of $P$, the transmission PSD. When $P$ is sufficiently low, the optimal route is the same as the cooperative one. That is, each relay node passes the message to the adjacent relay node that is closer to the destination. As $P$ increases, relay nodes instead pass the message to relays further down the line towards the destination. In fact, when $P$ is sufficiently large, the optimum (i.e., $\tau_{nc}$ minimizing) strategy is for the source to transmit directly to the destination.

The cooperative gain, defined as the ratio $\tau_{nc}/\tau_c$, is plotted in Fig. 1 for unit-spaced nodes ($D = 100$, $N = 100$, $B = 20$ nats) as a function of the system-wide transmission power $PW_T$. The curve is piece-wise linear. The non-differentiable break points correspond to the powers at which the optimum non-cooperative (shortest-path) route changes. For example, for (roughly) $0 \leq PW_T \leq 8$ all 100 nodes participate, for $8 \leq PW_T \leq 24$ half the nodes participate, for $24 \leq PW_T \leq$ 47 one-third participate, for $47 \leq PW_T \leq 78$ one-quarter participate and so forth.

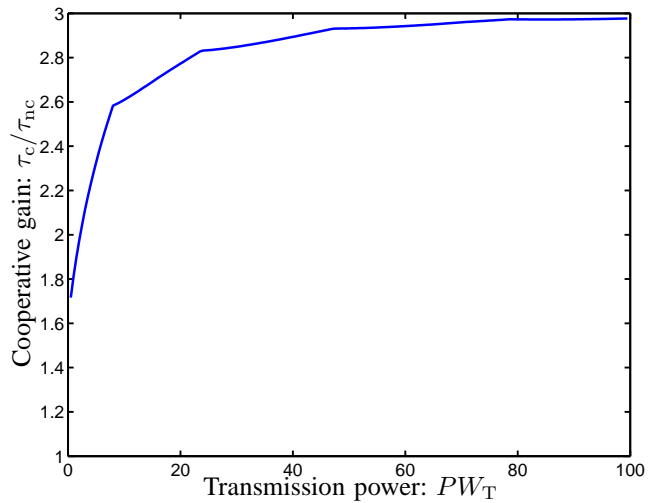

Fig. 1. Cooperative Gain of the One Dimensional Network.

As $N$ approaches infinity, and $P$ approaches zero, so that the product $PN^2$ stays small, we show in Section IV-E that the cooperative gain converges to $\pi^2/6 \simeq 1.64$. As can be seen by inspecting Fig. 1, the cooperative gain is greater at higher transmission PSDs.

Note that in this example since $P_i = P$ for all $i$ and the sum-bandwidth is fixed, the energy expended by the cooperative and non-cooperative schemes is $\tau_c P W_T$ and $\tau_{nc} P W_T$, respectively. In this case the ratio $\tau_c/\tau_{nc}$ is the same as the ratio between the energy expended in the cooperative and non-cooperative cases. We subsequently show that this is a general characteristic of equal-transmit PSD sum-bandwidth constrained system.

While the example one-dimensional network has an extremely simple topology, it illustrates two central characteristics of routing with mutual-information accumulation. First, the use of mutual-information accumulation decreases packet latency and energy usage. Second, the optimum route in a network with mutual-information accumulation can be quite different from the optimum route in a multihop network. These characteristics carry over to more complicated (and more practically relevant) two-dimensional networks. Illustrative examples of two-dimensional networks are given in Sec. VI.

## IV. Centralized Algorithms

We now consider the general task of optimizing route and resource allocations for two-dimensional networks with arbitrary attenuation between nodes. Our strategy is first to introduce the idea of the "transmission order".[2] This is the order in which the nodes are allowed to come on-line as transmitters. We can think of the transmission order as the route used by the cooperative scheme. Since a node cannot transmit until it has decoded the message, a node's position in the transmission order put constraints on the resources allocated to transmitters prior to it in the order. We then iterate between two sub-problems:

---

[2]In earlier papers, [14], [15] in the place of "transmission order " we used the term "decoding order".

1) First, for the given transmission order, we determine the optimum transmission parameters. This resource allocation problem turns out to be a linear program (LP).
2) Second, based on the solution of the LP we revise the transmission order.

In Sec. IV-A we provide a parametrization of the routing problem and show that, given a particular transmission order, the resource allocation problem can be expressed as an LP. In Section IV-B we show how to use the solution of the LP to generate a new transmission order that is at least as good in terms of end-to-end delay as the previous order. As indicated above, our final route and resource allocation algorithm, presented in Sec. IV-C, iterates between (a) solving an LP to find the optimal allocations for the current transmission order, and (b) revising the order to get an order with a lower delay. This iterative procedure finds a very good locally optimal (and often globally optimal – as we have verified on small networks) route and the corresponding resource allocations efficiently, even for very large networks.

*A. Problem parametrization and LP-based resource allocation*

Our parametrization of the routing problem revolves around the "transmission order". We define the transmission order by starting with any ordering of the $N+1$ network nodes where the source node is the first node in the order. The transmission order is the sub-sequence that starts with the source node, always labelled 0, and ends with the destination node, always labelled $L$ where $1 \leq L \leq N$. The transmission order indicates the order in which nodes can come on-line as transmitters. Since each node must decode before it can transmit, a node's position in the order puts constraints on the mutual information that that node must accumulate from earlier nodes in the order. As nodes $L+1, \ldots N$ never transmit (since they come on-line after the destination decodes), they are not considered part of the transmission order.

We denote the time at which node $i$ decodes the message as $T_i$ where $T_0 = 0$ and $T_L$ is the duration of the source-to-destination transmission. Instead of working with the $T_i$ we find it more useful to work with the inter-node decoding delays, $\Delta_i$, where $\Delta_i = T_i - T_{i-1}$ for $1 \leq i \leq L$. Message transmission can be thought of as consisting of $L$ phases. The $i$th phase is of duration $\Delta_i$ and is characterized by the fact that at the *end* of the phase the first $i$ nodes have all decoded the message.[3] We refer to each phase as a "time-slot". Time-slots are not of pre-set or equal lengths, rather their lengths are solved for in the optimization problem stated next.

For a given transmission order we find the resource allocation minimizing end-to-end delay $T_L$. The objective function is

$$T_L = \sum_{i=1}^{L} \Delta_i. \quad (3)$$

We minimize this linear objective function subject to the following constraints: (i) $\Delta_i \geq 0$ for all $i$, (ii) node $i$ must decode by time $T_i = \sum_{l=1}^{i} \Delta_l$, (iii) the energy constraint(s), and (iv) the constraint(s) on the use of time and bandwidth. We state constraints (ii)–(iv) in turn.

[3]In fact, as will become more clear when we discuss finding the best transmission order, additional nodes may have already decoded. But the first $i$ node are guaranteed to have already decoded.

First consider the decoding constraints. We express each of the $L$ such constraints as

$$\sum_{i=0}^{k-1} \sum_{j=i+1}^{k} A_{i,j} C_{i,k} \geq B \quad \text{for all} \quad k \in \{1, 2, \ldots, L\} \quad (4)$$

where

$$A_{i,j} \geq 0 \quad \text{for all} \quad i \in \{0, 1, \ldots, L-1\}, \ j \in \{1, 2, \ldots, L\}.$$

The $A_{i,j}$ are the degrees-of-freedom, i.e., the time-bandwidth product (or "area" in sec-Hz) used by the $i$th node in the $j$th time slot. Recall that $C_{i,k}$ is the spectral efficiency (bits/sec/Hz) of the channel connecting the $i$th transmitter to the $k$th receiver. The $k$th node is required (by definition) to decode by the end of the $k$th time slot. Eq. (4) says that the total mutual information flow to the $k$th node must exceed $B$ bits by the end of the $k$th time slot. Only the first $k-1$ nodes, that are earlier in the transmission order, can contribute to this sum.

The constraints (4) only include nodes in the transmission order. Not all $N+1$ nodes in the network need be included. For instance, if one node (neither source nor destination) is far from the rest (or masked by a building), then including decoding constraints for it in the set (4) would increase the total delay $T_L$. As we discuss when we present the "swapping" algorithm that improves the transmission order, nodes can be swapped out of the order. Such nodes are then no longer treated as part of the network and $L$ is decreased by one.

Next we consider constraints on energy and bandwidth. These can take the form of either *sum* constraints that apply to the sum-allocation across all nodes or *per-node* constraints that are applied to each node individually. Either type of energy constraint can be paired with either type of bandwidth constraint. Alternately, both sum and per-node constraints can be enforced. In the following subsections we describe the specifics of each case.

*1) Sum-energy constraint:* A sum-energy constraint $E_T$ is expressed as

$$\sum_{i=0}^{L-1} \sum_{j=1}^{L} A_{i,j} P_i = \sum_{i=0}^{L-1} \sum_{j=i+1}^{L} A_{i,j} P_i \leq E_T. \quad (5)$$

where the equality holds because $A_{i,j} = 0$ for $j \leq i$. This is true since node $i$ has not decoded until the end of slot $i$ and therefore can only transmit (and therefore would only be allocated positive bandwidth) in slots $i+1, \ldots, L$.

*2) Per-node energy constraint:* For the case of per-node energy constraints $E_i$ we replace (5) with

$$\sum_{j=i+1}^{L} A_{i,j} P_i \leq E_i \quad \text{for all} \quad i \in \{1, 2, \ldots, L\}. \quad (6)$$

*3) Sum-bandwidth constraint:* A sum-bandwidth constraint $W_T$ applied across all nodes can be expressed in terms of the time-bandwidth product allocated to each user in each time slot as

$$\sum_{i=0}^{j-1} A_{i,j} \leq \Delta_j W_T \quad \text{for all} \quad j \in \{1, 2, \ldots, L\}. \quad (7)$$

*4) Per-node bandwidth constraint:* If system bandwidth is divided into parallel channels, which can be allocated at most a single transmitter at any given time, we need impose bandwidth constraints on a per-node basis. In this case, instead of the $L$ constraints in (7), this model results in $L^2$ constraints, one per node per time slot:

$$A_{i,j} \leq \Delta_j W_i \quad \text{for all} \quad \begin{array}{l} i \in \{0,1,\ldots L-1\} \\ j \in \{1,2,\ldots,L\} \end{array}. \quad (8)$$

Commonly, each parallel channel may be of the same bandwidth so that $W_i = W_{\text{node}}$ for all $i$.

*5) Discussion of bandwidth constraints:* We now make some comments respecting the sum-bandwidth and per-node bandwidth constraints. Considering the sum-bandwidth constraint, several aspects of (7) are worth noting. First, the specific time-bandwidth allocation to each node *within* each transmission slot is not specified. This is because we model the fading as block-fading and frequency-flat. Therefore, within the transmission band, each transmitter is agnostic as to what is its exact time-bandwidth allocation. Degrees-of-freedom are treated like a fluid, and only the allocated time-bandwidth product is important. Our ideal rateless codes (and associated modulation techniques) are assumed to be able to use optimally whatever region of the spectrum is allocated each node for transmission.

Because the degrees-of-freedom are treated as a fluid, the optimal solution under a sum-bandwidth constraint can always be implemented by scheduling just one node to transmit at any given instant. In time slot $j$ we allocate the whole bandwidth to node $i$ for duration of $A_{i,j}/W_T$ sec. The ordering of transmissions within a time slot is immaterial since only at the end of the time slot do we require the next node in the order to be able to decode.

When both sum-energy and sum-bandwidth constraints are applied, we have the following theorem, proved in Appendix A.

**Theorem 1.** *Under a sum-bandwidth constraints, if $P_i = P$ for all $i$ then the solution that minimizes delay also minimizes the sum energy.*

This theorem tells us that in this setting there is no trade off between energy and delay. The minimum-energy route is identical to the minimum-delay route. We give an example in Section VI.

Per-node bandwidth and transmission PSD constraints are useful for modeling, e.g., ultra-wideband communication systems. In ultra-wideband systems, available bandwidth and transmit power are determined by frequency regulators [26]. Furthermore, constraints on the spreading factor are imposed by limits on hardware complexity as well as requirements of communications standards [27]. Consequently, a large number of orthogonal channels can be available, with each node being able to use exactly one of them.

*6) Alternate Objective Functions:* The LP framework presented is flexible enough to accommodate a number of useful objective functions that can take the place of delay. For example, instead of delay minimization one might rather minimize the sum-energy expenditure

$$\sum_{i=0}^{L-1} \sum_{j=i+1}^{L} A_{i,j} P_i$$

subject to an end-to-end delay constraints $\sum_{i=1}^{L} \Delta_i \leq \tau_{tot}$, as well as bandwidth constraints.

Alternately, one might be interested in minimizing the time-bandwidth footprint. This could be used to improve the performance of parallel transmissions (between different source-destination pairs) within the network under consideration, or could be used to minimize inter-network interference (if multiple networks are operating in the same area). In this case one could choose the objective function to be

$$\sum_{i=0}^{L-1} \sum_{j=1}^{L} A_{i,j}$$

subject to delay and energy constraints.

Finally, in the place of the unicast setting on which we focus in this paper, multicasting can also be addressed in the current framework by appropriately adjusting the objective function and constraints. We discuss the multicasting scenario further in Section IV-D.

*B. Optimizing transmission order*

The use of mutual information accumulation makes the optimum transmission order quite different from the non-cooperative multi-hop route. Because the accumulation of mutual information by each node extends across many time slots, the decoding process can have a very long memory. This makes it impossible to solve for the best transmission order efficiently through dynamic programming. At the same time since in a network of $N+1$ nodes there are $\sum_{i=0}^{N} \frac{(N-1)!}{(N-1-i)!}$ distinct orderings ($> 10^{63}$ for $N = 50$), exhaustive search of all orderings quickly exceeds computational capabilities.

In this section, we present a theorem that tells us how to improve the transmission order by exploiting the characteristics of the LP solution obtained in Section IV-A. Consider an arbitrary transmission order. Define

$$\mathbf{x}^* = \left[\Delta_1^*, \ldots, \Delta_L^*, A_{0,1}^*, A_{0,2}^*, \ldots A_{0,L}^*, A_{1,2}^*, \ldots, A_{L-1,L}^*\right]$$

to be the optimum solution obtained by the linear program for the order. Denote the optimum decoding delay as $T_L^* = \sum_{i=1}^{L} \Delta_i^*$. The following theorem is proved in Appendix B.

**Theorem 2.** *If $\Delta_i^* = 0$, use $T_L^{**}$ to denote the optimum decoding delay (under the same energy and bandwidth constraints) of the "swapped" transmission order:*

$$\begin{array}{ll} [0,\ldots,i-2,i,i-1,i+1,\ldots,L] & \text{if} \quad i \leq L-1 \\ [0,\ldots,L-2,L] & \text{if} \quad i = L \end{array}. \quad (9)$$

*Then $T_L^{**} \leq T_L^*$.*

The intuition behind Theorem 2 is illustrated in Figure 2. A solution to the LP with $\Delta_i = 0$ indicates that either node $i$ decodes at exactly the same time as node $i-1$ (which will never be the case in reality) or that, although later in the order, node $i$ can actually decode before node $i-1$. Therefore, swapping the positions of nodes $i$ and $i-1$ in the order will typically gives a decrease in the $T_L$ once the LP is solved for the revised order. If $i = L$ the destination is swapped with the node prior to it in the order. In this case that node $(L-1)$ is dropped from the order.

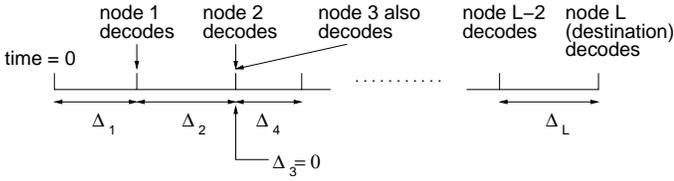

Fig. 2. Intuition behind order-swapping algorithm for $\Delta_3 = 0$.

*C. Algorithms for route & resource allocation optimization*

We are now in position to state the iterative route optimization algorithm. The algorithm alternates between revising the decoding order and solving the resulting LP until a route with locally optimal delay is obtained. While in general we obtain a local minimum, for small networks (of, e.g., 15 nodes, where we can exhaustively search all orders) we almost always reach the global optimum. Additionally, since the algorithm is quite efficient, we can try a number of different initializations to avoid particularly bad local minima.

**Algorithm 1:**
1) Start with an initial transmission order.
2) Use the linear program of Section IV-A to solve for the parameters of the minimum-delay solution.
3) Based on Theorem 2 adapt the transmission order to find an ordering whose minimum-delay solution is upper bounded by the delay of the current solution. Specifically:
   a) For any $i$ such that (a) $\Delta_i = 0$ and (b) $\Delta_{i-1} \neq 0$, swap the positions of the two nodes in the transmission order.
   b) If the node $L-1$ is swapped with node $L$, drop (the former) node $L-1$ from the order entirely. The resulting order contains only $L-1$ nodes.
4) Repeat steps 2)–3) until an ordering is obtained with an associated set of parameters $\mathbf{x}^*$ satisfying $\Delta_i^* > 0$ for all $i$. At this point terminate the algorithm.

Since the number of constraints in the linear program is linear in the network size, and the swapping algorithm is very simple, the routing algorithms can be applied to quite large networks.

In the following sub-sections we discuss various aspects of the algorithm in more depth, such as initialization and characteristics of certain special cases.

*1) Initialization:* If we initialize Algorithm 1 with an arbitrary transmission order at the target energy constraint(s) we typically find that $\Delta_i^* = 0$ for too many nodes for the search of the order space to get started. To address this issue we introduce the following algorithm that starts from feasible transmission order and (perhaps) relaxed energy constraint corresponding to that order. Following the presentation of Algorithm 2 we specify the choices we make in various cases.

**Algorithm 2:**
1) Initialize the algorithm with the initial transmission order and corresponding energy constraint.
2) Tighten the energy constraint slightly.
3) Use Algorithm 1 to re-optimize the route under the new energy constraints.
4) If the energy constraint now equals the target energy, terminate the algorithm. Otherwise, using the newly found route, return to step 2).

Algorithm 2 solves a sequence of route optimizations using Algorithm 1 under tighter and tighter energy constraints until the target energy is met. The optimized route found under one energy constraint is used to initialize Algorithm 1 under the next, slightly tighter, energy constraint. As with most non-linear iterative optimization routines, the choice of step size is important. In Algorithm 2 the step size corresponds to the increment by which the energy constraints are tightened. Ideally, the energy constraints are tightened only enough that a single $\Delta_i^* = 0$. This can typically be accomplished by making the increment small or dynamically choosing the increment. That is, if the energy constraint is tightened too much (multiple $\Delta_i = 0$), one can reduce the increment and re-optimize.

We now discuss the initial transmission order we use for specific cases. When per-node bandwidth constraints (8) are applied we initialize Algorithm 2 with the "flooding" order, while when a sum-bandwidth constraint is applied (7) we instead initialize using the non-cooperative route found via Dijkstra's shortest-path algorithm. First, consider per-node bandwidth constraints. In this setting there is a trade-off between energy and delay. At one extreme, when the energy constraint is fully relaxed, nodes are allowed unlimited energy consumption and the network can thereby achieve the minimum possible transmission delay. The transmission order at this extreme is what we term the flooding order, which is easily found as follows. The source node starts transmitting at time 0. Other nodes join in and begin transmitting as soon as they decode. All nodes continue to transmit until the destination decodes. The flooding order and corresponding energy can then be used to initialize Algorithm 2.

In contrast, when a sum-bandwidth constraint is imposed the flooding order cannot be used to initialize the system. This is because whenever a new node come on-line in the flooding order the bandwidth used increases and the sum-bandwidth constraint may be violated. Instead, for these networks we construct our initial transmission order starting from the non-cooperative shortest-path route. If nodes do not perform mutual information accumulation, and if nodes only receive in the time-slot immediately preceding the time at which they decode, then it is easy to solve for the optimum such non-cooperative path using the Dijkstra Algorithm [22]. As our initial transmission order we add to this shortest-path route the nodes that are able to decode the packet when non-cooperative shortest-path routing is used and all other nodes use mutual information accumulation. We calculate the energy used by this route and initialize the energy constraint accordingly.

*2) Characteristics of final route:* The mechanism that keeps our algorithm from necessarily reaching the global optimum is the swapping of nodes out of the transmission order. That is, when the $L-1$th node is swapped with node $L$ (the destination), it no longer enters the LP formulation. In particular this makes the decoding constraint (4) easier to meet. Intuitively, enforcing that nodes that are located further from the source than is the destination be able to decode via (4) can significantly increase the objective (the end-to-end transmission duration). However, it may turn out that a node that was swapped out of the transmission order could have ultimately prove useful. Our algorithm does not reintroduce nodes and so can converge to a sub-optimal solution.

Because of the exponential number of orderings we expect the problem of finding the optimal transmission order to be NP-hard. Note that for a special case of our problem, namely the low SNR limit where mutual information accumulation and energy accumulation become identical, Maric and Yates [10], [11] already proved that finding the optimal route is NP-hard. Thus, it is not surprising that there must be a caveat to our algorithm. However, our empirical observation is that, as long as the solution space is "smooth", as one reduces the energy from that used to initialize the search, one almost always reaches the global optimum. On the other hand, we have also constructed networks where at high energy one route is optimal, and at low energy a very different route is optimal, requiring the participation of nodes that do not decode at higher energies and therefore our algorithm drops from the transmission order. This might occur, for example, when the two routes are practically disconnected from one another by the shadowing of a large building.

Here is a simple such example consisting of four nodes. Node 0 is the source and node 3 is the destination. Each node has the same transmit power, $P_i = 1$ for all $i$, and each node is assigned a unit-bandwidth individual frequency band, i.e., equal per-node bandwidth constraints $W_i = 1$ for all $i$. Consider the situation where $B = 1$, $W_{\text{node}}C_{0,1} = 7\,\text{bits/sec}$, $W_{\text{node}}C_{0,2} = 5\,\text{bits/sec}$, $W_{\text{node}}C_{0,3} = 4\,\text{bits/sec}$, $W_{\text{node}}C_{1,2} = 0\,\text{bits/sec}$, $W_{\text{node}}C_{1,3} = 4\,\text{bits/sec}$, and $W_{\text{node}}C_{2,3} = 17\,\text{bits/sec}$. When the system has no energy constraint, the flooding order is $[0, 1, 3]$. Node 1 decodes at $1/7$ second. Then both source and node 1 transmit for another $3/56$ second, and the destination then decodes. The transmission duration is $\frac{11}{56} \simeq 0.196$ seconds and the energy consumption is $\frac{1}{7} + 2\frac{3}{56} = 0.25$. Node 2 never decodes. On the other hand, the minimum energy order is $[0, 2, 3]$. Node 2 decodes at $1/5$ second. The source turns off and node 2 starts transmitting. The destination decodes $(1 - 4/5)/17$ seconds later. Node 1 never decodes. The transmission duration is $\frac{18}{85} \simeq 0.21$ seconds and energy consumption is also 0.21 since only one node transmits at a time. In contrast, if either only the source transmits, or the source transmits until node 1 decodes and then node 1 transmits by itself until the destination decodes, the transmission duration is 0.25 seconds and the energy consumption is 0.25. In both these cases the energy consumption is identical to the flooding route (though the peak bandwidth use is one channel compared to the two channels used when the source node and node 1 transmit concurrently in the flooding route). Thus, without a way to re-introduce node 2 into the transmission order our algorithm would not obtain the optimum minimum energy solution when initialized with the flooding order.

One can consider heuristics for re-introducing nodes into the decoding order. For example, one might query nodes that have been dropped from the transmission order about whether they can decode at the current solution, and if they can, reintroduce them into the transmission order. One can see from the four-node example above that since node 2 doesn't decode when the flooding order is used, use of this particular heuristic does not necessarily result in the optimum minimum-energy route being found.

### D. Multicasting

The basic multicasting scenario (sending a common message to all nodes) requires all nodes to decode. The only change required in the various versions of the LP stated in (4)–(7) to yield a multicast solution is that $L$ becomes $N$.

In contrast to the unicasting, in multicasting nodes are never dropped from the transmission order. The main cause for our algorithm only achieving local rather than global optimality discussed in Sec. IV-C2 is thereby obviated. Therefore, we should nearly always achieve the global optimum using our iterative approach. The one remaining caveat is the step-size; it is important to reduce the energy constraint between LPs in small enough increments that only one $\Delta_i$ goes to zero per iteration. In a real-world network this will be the case, but in an artificial network it is possible to coordinate node-to-node gains $h_{i,j}$ so that multiple $\Delta_i$ go to zero at the same time.

There is also a multicasting problem between unicasting and basic multicasting where we require some subset of the $N+1$ nodes to decode. This scenario is also easy to incorporate into our framework. One simply never drops any of these (now multiple) "destination nodes" from the transmission order. In term of the LP, node $L$ is the index of the last of these destinations to decode.

### E. One-dimensional networks

In this section we derive our results for simple one-dimensional networks under constant PSD $P_i = P$ for all $i$. To recap the discussion of Sec. III, we assume that there are $L+1$ nodes equally spaced along the line segment $[0, D]$ with path-loss that decays quadratically with distance. End-to-end delay is be minimized under a sum-bandwidth constraint. The topology and monotonic path-loss imply that the minimum energy transmission order is $[0, 1, \ldots, L-1, L]$. Furthermore, the sum-bandwidth constraint implies that only one node is active per time-slot – the node closest to the destination that has decoded. The source node only transmits in time slot 1, the first node only in time slot 2 and the $i$th node only in time slot $i + 1$. The transmission delay can then be immediately computed through equations $A_{0,1}C_{0,1} = B$, $A_{1,2}C_{1,2} + A_{0,1}C_{0,2} = B$, and in general

$$\sum_{i=1}^{k} A_{i,i+1} C_{i,k} = B \qquad (10)$$

for each $k$ such that $1 \leq k \leq N - 1$. Since $C_{0,i} = C_{j,i+j}$ we can write the equations (10) in the matrix form as

$$\begin{bmatrix} C_{0,1} & 0 & \cdots & 0 \\ C_{0,2} & C_{1,2} & \ddots & \vdots \\ \vdots & \vdots & \ddots & 0 \\ C_{0,N} & C_{1,N} & \cdots & C_{N-1,N} \end{bmatrix} \begin{bmatrix} A_{0,1} \\ A_{1,2} \\ \vdots \\ A_{N-1,N} \end{bmatrix} = \begin{bmatrix} B \\ B \\ \vdots \\ B \end{bmatrix}.$$

Let $\mathcal{K}$ denote the lower triangular matrix containing the $C_{i,k}$. As the length of the $i$th time slot is $A_{i-1,i}/W_T$, the transmission delay $\tau_c$ can be calculated as

$$\tau_c = \frac{\sum_{i=1}^{N} A_{i-1,i}}{W_T} = \frac{B}{W_T} \times [1 \ldots 1] \mathcal{K}^{-1} \begin{bmatrix} 1 \\ \vdots \\ 1 \end{bmatrix}.$$

Since $P_i = P$ for all $i$ we know by Theorem 1 that the minimum delay route is also the minimum energy route. This result is especially apparent for this network. The node closest to the destination that has already decoded also has the best channels to all remaining nodes that have not yet decoded. When $P_i = P$ for all nodes, it also has the highest information flow $C_{i,k}$ to those remaining nodes. Thus, not only should that node transmit but, under a sum-bandwidth constraint, it should be allocated all the bandwidth. Energy is therefore not expended anywhere else and the minimum energy and minimum delay routes are the same. Even if node PSDs are not all the same, the optimum decoding order remains the same because of the linear topology of the network. The linear program can then be solved to find the optimum $\{A_{i,j}\}$. One should note that when the $P_i$ are not all the same, there may be an energy-delay trade off, even for this simple linear network.

When there are a large number of nodes $N$ and when $P$ is small, the cooperative gain $\tau_{\rm nc}/\tau_{\rm c}$ takes on a particularly simple form. By $N$ large and $P$ small we mean that the product $N^2 P$ is small. Under this assumption the spectral efficiency

$$C_{i,k} = \log_2\left[1 + \frac{h_{i,k}P}{N_0}\right] = \log_2\left[1 + \frac{N^2}{(k-i)^2 D^2}\frac{P}{N_0}\right]$$

between any two nodes is well approximated as $\log_2 e \; \frac{N^2}{(k-i)^2 D^2}\frac{P}{N_0}$. As mentioned in Section III, when $P$ is small, the shortest path route for the non-cooperative scheme is the same as for the cooperative scheme – multi-hop through every node. We term the incremental decoding delay incurred by each node in this route $\Delta\tau_{\rm nc}$ where the overall delay is $\tau_{\rm nc} = N\Delta\tau_{\rm nc}$. The incremental delay is calculated as $B = C_{j-1,j}W_{\rm T}\Delta\tau_{\rm nc} \simeq \log_2 e \frac{P}{N_0}\frac{N^2}{D^2}W_{\rm T}\Delta\tau_{\rm nc}$, and solving for $\Delta\tau_{\rm nc}$ gives

$$\Delta\tau_{\rm nc} = \frac{1}{\log_2 e}\frac{BN_0}{PW_{\rm T}}\frac{D^2}{N^2}.$$

When nodes accumulate mutual information the incremental delay is reduced. The decoding constraint of the $k$th node is $B = \sum_{l=1}^{k} C_{k-l,k} A_{k-1,k-l+1}$. In a large network ($N$ large) the $A_{j,j+1}$ will approach a steady state value for $j \gg 0$. The length of each time-slot will also approach a steady state value $\Delta\tau_{\rm c}$. For such $j$, since the node is allocated all bandwidth for duration $\Delta\tau_{\rm c}$, the corresponding allocation $A_{j,j+1} = \Delta\tau_{\rm c}W_{\rm T}$. In the asymptotic limit of $N$ large these time-slots dominate the overall delay. In this regime we calculate $\Delta\tau_{\rm c}$ as $B = \sum_{l=1}^{k} C_{k-l,k}W_{\rm T}\Delta\tau_{\rm c} = W_{\rm T}\Delta\tau_{\rm c}\log_2 e \frac{PN^2}{N_0 D^2}\sum_{l=1}^{k}\frac{1}{l^2}$. Letting $N$ (and $k$) go to infinity, we have $\sum_{l=1}^{\infty}\frac{1}{l^2} = \frac{\pi^2}{6}$, giving in the limit

$$\Delta\tau_{\rm c} = \frac{1}{\log_2 e}\frac{BN_0}{PW_{\rm T}}\frac{D^2}{N^2}\frac{6}{\pi^2}.$$

The cooperative gain is then calculated as

$$\frac{\tau_{\rm nc}}{\tau_{\rm c}} = \frac{N\Delta\tau_{\rm nc}}{N\Delta\tau_{\rm c}} = \frac{\pi^2}{6}.$$

## V. DISTRIBUTED ALGORITHMS

It is often not desirable or even possible to centralize the routing routine. In centralized solutions all channel state information (CSI) must be aggregated centrally. The resulting routing information is then dispersed throughout the system. Limitations on centralized solutions are particularly constraining in the following circumstances:

- *Large networks*: Since the number of possible links (and thus CSI that has to be distributed) increases as $(L+1)!$, aggregating the CSI of all links can incur an unacceptable overhead if $L$ is large.
- *Temporally varying networks*: Even in small networks time-slotting and other restrictions can cause the CSI to be outdated by the time it arrives at the central location.

To address these issues we describe two distributed algorithms inspired by the characteristics of our centralized solution. These algorithms require far less CSI, perform mutual information accumulation, and yield performance nearly as good as the centralized algorithms.

### A. Distributed Algorithm 1

Our first distributed algorithm commences with a direct transmission from source to destination. In an iterative fashion intermediate nodes are added to the route.[4] Specifically, the source transmits a sounding signal. All nodes estimate their channel from the source. The destination replies with a second sounding signal. Nodes then estimate their channel to the destination. Given this pair of CSI measurements each node determines the potential energy savings if it were to join the path. Potential energy savings are calculated as

$$\frac{B}{W_{\rm T}}\frac{(C_{i,L} - C_{0,L})(C_{0,i} - C_{0,L})}{C_{0,i}C_{0,L}C_{i,L}}.$$

Each node then broadcasts this information to the rest of the network using any of the many available contention multiple access schemes. The node with the highest energy saving is chosen to participate. In the next step, the CSI from that node to all other nodes in the network is determined. Again, all nodes analyze whether they can save energy by joining the route. The process continues until no further energy savings are possible.

The algorithm is simple and, as we see in Sec. VI, very effective. It does has one drawback. The initial setup of a route takes a long time. This is because the starting point of the algorithm is a direct source-to-destination transmission. If the source-to-destination pathloss is high, a long sounding signal is required (noise averaging over a long time results in a good estimate of the channel strength). Adding nodes progressively shortens the transmission delay. Once a route is set up, changes (due to changing channel conditions) can be done rather efficiently, since the route can be modified without tearing down and rebuilding it from scratch.

### B. Distributed Algorithm 2

A somewhat simpler distributed algorithm can be implemented as follows. The destination broadcasts a sounding signal and all nodes estimate their channels to the destination. Each node broadcasts its own node-to-destination CSI to all other nodes. The source then starts to transmit the information packet. The first node that can decode the data and has a better channel to the destination then takes over and the source node

---
[4]The principle of the algorithm is somewhat similar to the PAR algorithm described in [23].

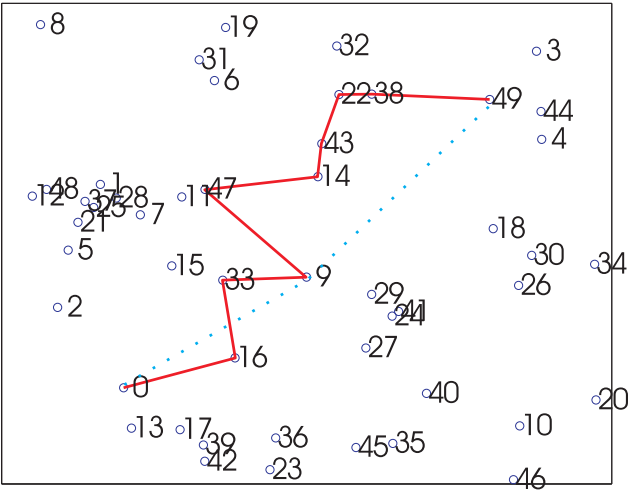

Fig. 3. Location of nodes in fifty node network. The minimum-energy cooperative routing is shown.

turns off. New nodes continue to replace previous nodes until the message reaches the destination.

Because of the lack of full, network-wide, CSI, the two algorithms presented in this section require the use of rateless codes. This is in contrast to the centralized algorithms which can use block codes as the length of each time slot is known apriori. As mentioned in Section II, however, while mutual information accumulation can theoretically be implemented using generic block codes, the particular structure of rateless codes makes it much easier to implement.

## VI. NUMERICAL DETAILS OF RESULTS

In this section we give detailed numerical results for the algorithms developed in this paper under various constraints. These results further exemplify the basic features of routing with mutual information accumulation described in the discussion of one-dimensional networks in Sec. III.

Our examples concern two-dimensional networks located in the unit square. For all examples the source node 0 is located at $[0.2, 0.2]$ and the destination node 49 is located at $[0.8, 0.8]$. Remaining nodes are placed randomly according to the uniform distribution in the unit square. A typical wireless network from this ensemble is shown in Fig. 3. In order to give the reader a strong sense of the relationship between geometry and channel strength we study the case where the channel gain $h_{i,j}$ between node $i$ and node $j$ is deterministically related to the Euclidean distance $d_{i,j}$ between them as $h_{i,j} = (d_{i,j})^{-2}$.

To quantify the performance of our algorithm we establish a baseline non-cooperative strategy for comparison. For this comparison, we choose a multi-hop strategy. Only one node transmits at each time. The route is selected using Dijkstra's shortest path algorithm [22], and each node accumulates mutual information only from the node that immediately precedes it. We also consider a hybrid strategy that uses the Dijkstra-based route but where nodes perform mutual-information accumulation (listening to all previous transmission instead of just the immediately prior transmission). By studying both cases we get a sense of the fractional performance improvement due to the use of mutual information accumulation, and that due to using a route designed specifically for cooperative communication.

### A. System wide bandwidth constraint

We first consider a sum-bandwidth constraint on the specific network shown in Fig. 3 where $B = 28.9$ bits (20 nats), $N_0/2 = 1$, $W_T = 1$, and $P_i = P = 1$ for all $i$. Under sum-energy and sum-bandwidth constraints, as is proved in Thm. 1 the minimum-delay and minimum-energy routes are the same. Therefore, in this case there is no energy/delay trade off.

After solving for the route using our centralized algorithm, the subset of nodes that actually transmit in the final transmission order is $[0, 16, 33, 9, 47, 14, 43, 22, 38, 49]$, indicated in Fig. 3 by the solid line. As can be seen from inspection of the figure, the nodes that are active in the minimum delay (and therefore minimum energy) solution are the nodes that lie closest to the direct path between source and destination. This is due to the fact that channel gain is inversely proportional to distance squared. For this example network the destination decodes after $\tau_c = 13.09$ seconds.

We now develop results for a non-cooperative multihop routing example. In the non-cooperative case, and as described for linear networks in Section IV-E, the incremental delay accrued by the hop from node $i$ to node $j$ is $B/W_T C_{i,j} = B/W_T \log_2 \left[1 + \frac{h_{i,j} P}{N_0}\right]$. For the node placements in Fig. 3 the shortest path route is found to be $[0, 9, 49]$, indicated in the figure by the dotted line. The resulting source-to-destination delay $\tau_{\text{nc}}$ is 21.47 seconds. Interestingly, the set of nodes that transmit in the shortest path problem is a proper subset of those that transmit in the cooperative protocol. Furthermore, the only relay node participating in the optimal (shortest-path) route is the one closest to the direct path connecting source to destination.

The decrease in transmission duration obtained by our cooperative route compared to the non-cooperative approach stems from two causes: the use of mutual-information accumulation decoding and the use of a route tuned to cooperation. If the nodes perform mutual information accumulation, but only the nodes in route obtained from Dijkstra's algorithm participate in transmission, the transmission delay is 16.51 seconds. Thus, roughly half the decrease in transmission duration is due to the use of mutual information accumulation, and half due to the use of a route tuned to mutual information accumulation.

To ensure that the improvement is not specific to the sample network of Fig. 3, we calculate the distribution of decoding delays over an ensemble of 500 independently generated realizations of networks of the type depicted in Fig. 3 where the source and destination locations are held constant at $[0.2, 0.2]$ and $[0.8, 0.8]$, respectively, and the rest of the nodes are placed uniformly on the unit square.

The cumulative distribution function (CDF) of decoding delay is plotted in Fig. 4. The average delay of the *centralized* cooperative routing using mutual information accumulation is 12.54 seconds, while the average delay of non-cooperative routing, solved for using Dijkstra's *shortest-path* algorithm, is 21.52 seconds. On average, the conventional non-cooperative multihop transmission incurs additional delay and energy usage on the order of 70% as compared to cooperative transmission.

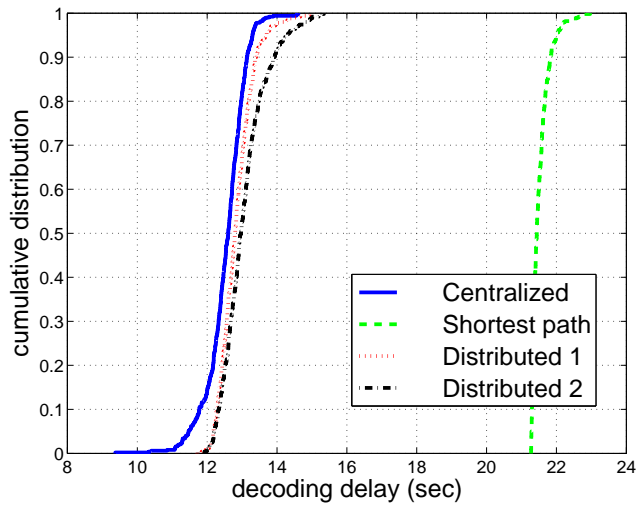

Fig. 4. Cumulative distribution of excess delay of distributed solutions as compared to centralized algorithm.

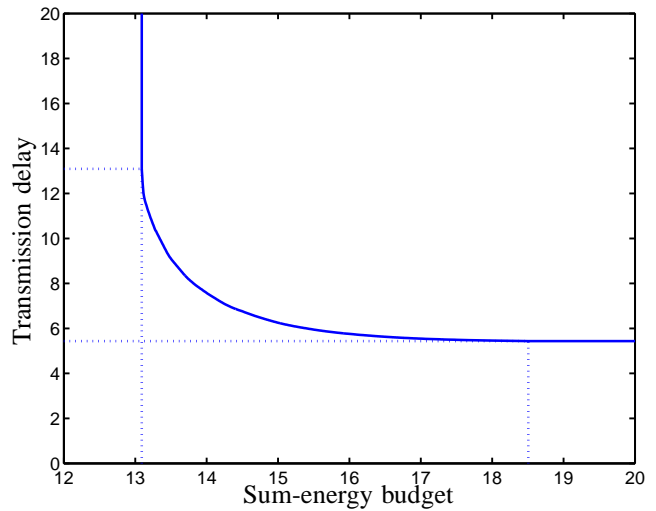

Fig. 5. Delay versus energy trade off in a fifty node network. Nodes are placed uniformly at random in the unit square. Channel gains between nodes separated by a distance $d$ are proportional to $d^{-2}$. The sum of energies over all nodes and the per-node bandwidth are limited.

In addition, on Fig. 4 we also plot CDF results for the two distributed routing algorithms introduced in Section V. The penalty for using the distributed algorithms in terms of delay (or, equivalently, energy) is small. On average the first distributed algorithm incurs less than $2.5\%$ excess delay as compared to the centralized solution. The excess delay of the second distributed algorithm is less than $4.2\%$. The distributed algorithms relax the need for centralized CSI at the cost of modest increases in delay.

### B. Per-node bandwidth constraint

In this section we again consider the network of Fig. 3, but this time under per-node bandwidth constraints. In this setting there is a trade off between system resources (energy and bandwidth) and transmission delay. We keep the same parameters as before, namely $B = 28.9$ bits (20 nats), $N_0/2 = 1$, $P_i = P = 1$, and we set the per-node bandwidth constraint $W_i = 1$ for all $i$. The energy-delay trade off achieved is plotted in Fig. 5.

At one resource extreme we flood the network, fully relaxing the sum-energy constraint and allowing nodes unlimited energy consumption. The network can then achieve the minimum possible transmission delay. In the network depicted in Fig. 3 all nodes except $3, 4,$ and $44$ participate in the flooding routing. The order in which nodes come on-line as transmitters is $[0, 13, 17, 39, 42, 16, 2, 36, 23, 15, \ldots, 20, 32, 34, 8, 49]$. The flooding energy is $18.5$ and the transmission delay is $5.4$.

As the energy budget is decreased, nodes with weaker connectivity to the destination go off-line and only nodes with stronger channels remain active. Finally, at some minimum energy, the network becomes disconnected. The limit point of delay as the energy approaches is defined as the minimum-energy transmission duration. For the network of Fig. 3 the minimum-energy route $[0, 16, 33, 9, 47, 14, 43, 22, 38, 49]$, depicted by the solid line. The minimum energy is $13.09$ and the minimum delay is $13.09$. The low-energy route has only a single transmitter transmitting at any given time. This is because if each node waits for all prior transmissions to complete before beginning its own transmission, that node will have accumulated the most mutual information possible. Therefore, the optimum route has only one node at a time transmitting. Since only one node at a time transmits, the system bandwidth is constant. And thus, in the low-energy limit the sum-bandwidth and per-node bandwidth constraints are fully comparable and, indeed, $\tau_\text{c} = 13.09$ for this network in the sum-bandwidth setting of Sec. VI-A. (Furthermore, since only one node at a time transmits and $P_i = 1$ for all nodes the minimum energy and minimum delay are identical).

When a larger energy budget is allowed, multiple nodes can transmit simultaneously. In contrast, when bandwidth constraints are imposed on a per-node basis, the non-cooperative scheme is limited to the transmission band of a single node. Therefore, the peak bandwidth used by the cooperative strategy when the transmission delay is minimized can exceed that of the non-cooperative strategy, though the total energy consumption will still be lower. For instance, for the example discussed in Sec. VI-A, $\tau_\text{nc} = 21.47$ and since $P_i = 1$ and $W_i = 1$ for all $i$, the energy consumption of the non-cooperative case is also $21.47$, which exceeds the cooperative flooding energy of $18.51$. Of course, for this case, the improvement of delay is more impressive: the flooding route has a delay of $5.4$ compared to the non-cooperative delay of $21.47$.

### VII. SUMMARY AND CONCLUSIONS

In this paper we analyze the problem of generalized routing in cooperative relay networks that use mutual-information accumulation. We split the routing problem into one of finding the best transmission order and one of finding the best resource allocation given a transmission order. As our solution is based on solving a sequence of linear programs, it is quite numerically efficient, even for large networks. We also show that under equal per-node PSDs, the minimum-delay solution also minimizes energy consumption. The resulting route is markedly different from the conventional shortest-path route. The delay (and energy usage) of the latter is about 70 % more in the examples we present. We also develop distributed algorithms that retain most of the performance gains without requiring centralized knowledge of channel state information.

The approach presented in this paper is a step towards practically realizing cooperative communications in large networks. Future work will focus on optimizing the power allocation (adjusting the $P_i$), algorithms that are suitable for imperfect channel state information, and the impact of non-ideal codes and hardware.

**Acknowledgements:** We thank Dr. Neelesh Mehta for useful discussions, and Dr. Jin Zhang, Dr. Kent Wittenburg, and Dr. Joseph Katz for their support and encouragement.

## APPENDIX

### A. Proof of Theorem 1

Start from the energy used $E_{\text{used}}$

$$E_{\text{used}} = \sum_{i=0}^{L-1} \sum_{j=1}^{L} A_{i,j} P_i = \sum_{i=0}^{L-1} \Delta_j W_{\text{T}} P = T_L W_{\text{T}} P. \quad (11)$$

Equality must hold in $(a)$ else (7) is loose at the optimum. But, this means that some degrees of freedom $A$ go unallocated in some times slot. If this is the case the decoding time can be strictly decreased by moving up all subsequent decoding times by $A/W_{\text{T}}$. Equality in $(b)$ holds by definition, $\sum_{i=0}^{L-1} \Delta_j = T_L$. Since the duration of decoding $T_L$ is proportional to the energy used $E_{\text{used}}$ minimizing one minimizes the other.

### B. Proof of Theorem 2

**Case 1:** ($i=1$) Combine node 1's decoding constraint (4) with the total degrees-of-freedom in time slot 1 (7) or (8), for the sum-bandwidth and per-node bandwidth constraints, respectively, to get

$$\frac{B}{C_{0,1}} \leq A_{0,1} \leq \Delta_1^* W_{\text{T}}. \quad (12)$$

for the sum-bandwidth constraint and

$$\frac{B}{C_{0,1}} \leq A_{0,1} \leq \Delta_1^* W_{\text{node}}. \quad (13)$$

for the per-node constraint. Equation (12) and (13) demonstrate for both cases the intuitive fact that no node can decode the message before the source. Therefore, $\Delta_1^* > 0$ is always true (for any ordering) and we need only consider $2 \leq i \leq L$.

**Case 2:** ($2 \leq i \leq L-1$) We show that $\tilde{\mathbf{x}}$, a "swapped" version of $\mathbf{x}^*$, is a feasible solution for the swapped ordering that has a decoding delay equal to the optimal decoding delay of the original ordering. Define

$$\tilde{\mathbf{x}} = \left[\tilde{\Delta}_1, \ldots, \tilde{\Delta}_L, \tilde{A}_{0,1}, \tilde{A}_{0,2}, \ldots \tilde{A}_{0,L}, \tilde{A}_{1,2}, \ldots, \tilde{A}_{L-1,L}\right],$$

where

$$\begin{aligned}
\tilde{\Delta}_i &= \Delta_i & \text{for all } i \\
\tilde{A}_{k,l} &= A_{k,l}^* & \text{for all } k,j \text{ s.t. } k \neq i-1, k \neq i \\
\tilde{A}_{i-1,i} &= 0 \\
\tilde{A}_{i-1,j} &= A_{i,j}^* & \text{for all } j \in \{i+1, \ldots, L\} \\
\tilde{A}_{i,j} &= A_{i-1,j}^* & \text{for all } j \in \{i+1, \ldots, L\}.
\end{aligned}$$

We immediately see $\sum_{i=1}^{L} \tilde{\Delta}_i = \sum_{i=1}^{L} \Delta_i^*$. We now show that $\tilde{\mathbf{x}}$ satisfies all problem constraints.

First note that the degree-of-freedom allocations $A_{i,j}$ made to each node in each time slot are almost all identical in $\mathbf{x}^*$ and $\tilde{\mathbf{x}}$. There are two exceptions. The first, $A_{i-1,i}$ doesn't appear in $\tilde{\mathbf{x}}$, but $A_{i-1,i} = 0$ since $\Delta_i = 0$. The second, $\tilde{A}_{i-1,i} = 0$.

From this we immediately get that the energy, decoding, and degrees-of-freedom constraints remain satisfied for $\tilde{\mathbf{x}}$.

First, since the non-zero degree-of-freedom allocations are identical for $\mathbf{x}^*$ and $\tilde{\mathbf{x}}$, the energy usage remains the same under either sum-energy or per-node-energy constraints. For the same reason the decoding ability of nodes $1, \ldots, i-2$, nodes $i+1, \ldots, L$, and the "old" (pre-swapped) node $i-1$ remain unchanged. The old node $i$ doesn't benefit from the old node $i-1$'s transmissions any longer since the order is swapped in $\tilde{\mathbf{x}}$. However, because $\Delta_i = 0$, $A_{i-1,i} = 0$ and it didn't accumulated any mutual information in the old order in any case. Finally, since the positive degree-of-freedom allocations remain the same, and the time-slot durations $\tilde{\Delta}_i$ remain the same, the degree-of-freedom constraints all remain satisfied.

**Case 3:** ($i = L$) For the same reasoning as in case 2, if we define the same vector $\tilde{\mathbf{x}}$, the decoding delay remains the same and all constraints remain satisfied. Now, if we drop the (new) node $L$ from the problem completely (the destination is the new node $L-1$) the reduced solution is still feasible since none of the other nodes relied on the dropped nodes transmission. (It was the last in the order).